\documentclass[pre,aps,10pt,superscriptaddress,twocolumn,floatfix]{revtex4-2}

\usepackage[nice]{nicefrac}
\usepackage{graphicx,color}
\usepackage{amsmath,amssymb,bm,bbm}
\usepackage{amsmath}
\usepackage{braket}
\usepackage{bbold}
\usepackage{float}
\usepackage{physics}

\usepackage[plainpages=false,pdfpagelabels,colorlinks=true,linkcolor=red,urlcolor=blue,citecolor=blue,pdftitle={},pdfauthor={},pdfdisplaydoctitle=true,pdfduplex=DuplexFlipLongEdge]{hyperref}

\definecolor{darkred}{rgb}{0.90,0.2,0.2}
\definecolor{darkgreen}{rgb}{0,0.60,.2}
\definecolor{darkblue}{rgb}{0.1,0.3,1}
\definecolor{grey}{cmyk}{0,0,0,0.25}
\definecolor{orange}{cmyk}{0,0.6,0.8,0}

\usepackage[T1]{fontenc}

\begin{document}
\title{Fading ergodicity meets maximal chaos}

\author{Rafał Świętek}
\affiliation{Department of Theoretical Physics, J. Stefan Institute, SI-1000 Ljubljana, Slovenia}
\affiliation{Department of Physics, Faculty of Mathematics and Physics, University of Ljubljana, SI-1000 Ljubljana, Slovenia\looseness=-1}
\author{Patrycja Łydżba}
\affiliation{Institute of Theoretical Physics, Faculty of Fundamental Problems of Technology, Wrocław University of Science and Technology, 50-370 Wrocław, Poland\looseness=-3}
\affiliation{Department of Theoretical Physics, J. Stefan Institute, SI-1000 Ljubljana, Slovenia}
\author{Lev Vidmar}
\affiliation{Department of Theoretical Physics, J. Stefan Institute, SI-1000 Ljubljana, Slovenia}
\affiliation{Department of Physics, Faculty of Mathematics and Physics, University of Ljubljana, SI-1000 Ljubljana, Slovenia\looseness=-1}

% \begin{bibunit}

\begin{abstract}
Fading ergodicity provides a theoretical framework for understanding deviations from the eigenstate thermalization hypothesis (ETH) near ergodicity-breaking transitions. In this work, we demonstrate that the breakdown of the ETH at the interaction-driven ergodicity-breaking critical point in the quantum sun model gives rise to to the maximally divergent fidelity susceptibility. We further extend our analysis to the energy-driven ergodicity-breaking transition associated with the many-body mobility edge. Specifically, we show that fidelity susceptibilities at energies away from the middle of the spectrum exhibit a divergent peak near the mobility edge.
Finally, we argue that fading ergodicity provides a simple and accurate description of the ETH breakdown in the quantum sun model, which is accompanied with the emergence of a peak in fidelity susceptibility and the onset of maximal chaos at the ergodicity-breaking critical point.
\end{abstract}

\maketitle

%%%%%%%%%%%%%%%%%%%%%%%%%%%%%%%%%%%%%%%%%%%%%%%%%%%%%%%%%%
\section{Introduction} \label{sec:introduction}

Quantum thermalization in isolated many-body systems, along with its potential breakdowns, constitutes a rich area of research that bridges statistical mechanics, condensed matter physics, and quantum information science, raising numerous unresolved questions~\cite{dalessio_kafri_16, mori_ikeda_18, sierant_lewenstein_25}.
The emergence of quantum thermalization is, among others, characterized by the agreement of Hamiltonian spectral properties with predictions of random matrix theory~\cite{bohigas_giannoni_84, casati_valzgris_80, montambaux_poilblanc_93, hsu_dauriac_93, distasio_zotos_95, prosen_99, santos_04, rabson_narozhny_04, rigol_santos_10}, and by the validity of the eigenstate thermalization hypothesis (ETH)~\cite{Deutsch91, srednicki_94, srednicki_99, rigol_dunjko_08}.
The latter is expressed as the ansatz for the matrix elements of observable $\hat V$ in eigenstates $|m\rangle$ of Hamiltonian $\hat H$, with $\hat H|m\rangle = E_m|m\rangle$,
\begin{equation} \label{def_eth}
    \langle n|\hat V|m\rangle = {\cal V}(\bar E)\delta_{nm} + \rho(\bar E)^{-1/2} f_{V}(\bar E,\omega) R_{nm} \;.
\end{equation}
We refer to Eq.~\eqref{def_eth} as the {\it conventional} ETH.
In the latter, $\rho(\bar E)$ is the density of states at energy $\bar E = (E_n+E_m)/2$ that scales as $\rho\propto 2^L$ in a system of $L$ qubits, ${\cal V}$ and $f_V$ are smooth functions of their arguments, $\omega=E_n-E_m$, and $R_{nm}$ is a random number with zero mean and unit variance.
It is understood that the ETH from Eq.~\eqref{def_eth} represents a sufficient condition for the emergence of quantum thermalization~\cite{dalessio_kafri_16}.

Another key focus of current research is the characterization of quantum phase transitions in the ground states of Hamiltonian systems~\cite{sachdevbook}.
A possible way to understand them is to consider the wavefunction fidelity~\cite{quan_song_06}, $F = \langle \psi_0(\lambda) | \psi_0(\lambda+\delta\lambda)\rangle$, where $|\psi_0(\lambda)\rangle$ is the ground state of Hamiltonian $\hat H(\lambda)=\hat{H}_0+\lambda\hat{V}$ with energy $E_{\psi_0}(\lambda)$, and we refer to $\lambda\hat{V}$ as a perturbation with strength $\lambda$.
At infinitesimally small perturbation, $\delta\lambda\to 0$, the variation of fidelity is characterized by the fidelity susceptibility~\cite{zanardi_giorda_07, you_gu_2007, gu_10},
\begin{equation}\label{eq:fidelity}
    \chi_0=\sum_{m\neq \psi_0}\frac{|\mel{\psi_0}{\hat V}{m}|^2}{\qty( E_m-E_{\psi_0} )^2}\;,
\end{equation}
hence it is determined by an interplay between the off-diagonal matrix elements of $\hat V$ and the properties of the level spacings $E_m-E_{\psi_0}$. In Eq.~\eqref{eq:fidelity}, we have made the dependence on $\lambda$ implicit.
Many studies reported evidence that the peak of $\chi_0$ versus $\lambda$ is a signature of a quantum phase transition, see, e.g., Refs.~\cite{gu_10, cozzini_giorda_07, cozzini_ionicioiu_07, buonsante_vezzani_07, zanardi_giorda_07, you_gu_2007, camposvenuti_zanardi_07, chen_wang_08, gu_kwok_08, jacobson_garnerone_09, capponi_schwandt_2009, capponi_albuquerque_2010, troyer_wang_2015, sierant_maksymov_19, wei_19}.

A novel perspective on the breakdown of quantum thermalization, which also offers connections to quantum phase transitions in ground states, was recently provided through the framework of adiabatic gauge potentials (AGPs)~\cite{demirplak_rice_03, demirplak_rice_05, berry_09, kolodrubetz_sels_17}.
In particular, the norm of AGPs was interpreted as a measure of sensitivity of all Hamiltonian eigenstates to perturbations, and hence a possible measure of quantum chaos~\cite{sels_pandey_2020}.
Then, the peak of the AGP norm, or of the fidelity susceptibility of excited eigenstates~\cite{Sels_2021, leblond_sels_21, skvortsov_amini_22, nandy_cadez_22, kim_polkovnikov_24, cadez_dietz_24}, can be understood as the onset of maximally chaotic behavior~\cite{lim_matriko_24}. 
The latter may be a hallmark of the ergodicity-breaking transition in finite systems~\cite{Sels_2021}.

This discussion raises several intriguing questions. Does the emergence of a peak in the AGP norm, or in the fidelity susceptibility of eigenstates far above the ground state, indicate the ergodicity-breaking critical point, analogous to the critical point of a ground-state quantum phase transition? Is the position of this peak correlated with the onset of the ETH breakdown? Furthermore, is there a common theoretical framework that simultaneously explains the suppression of the conventional ETH, as described by Eq.~\eqref{def_eth}, and the approach of the fidelity susceptibility of a typical excited eigenstate towards its maximal value?

Here we provide affirmative answers to the above questions.
We argue that the concept of fading ergodicity~\cite{kliczkowski_vidmar2024}, which describes a framework for the breakdown of quantum thermalization, naturally associates the breakdown of the conventional ETH from Eq.~\eqref{def_eth} with the emergence of a peak in the fidelity susceptibility and the AGP norm at the ergodicity-breaking critical point.

Our main results are two fold.
We first consider the interaction-driven ergodicity breaking transition, described by the quantum sun model~\cite{suntajs_vidmar_22,hopjan_vidmar_23a, hopjan_vidmar_23b, suntajs_hopjan_24}, and we show that the peak of the fidelity susceptibility agrees, to a high numerical precision, with the position of the ergodicity-breaking critical point in the thermodynamic limit, obtained from other measures.
This reaffirms the expectation~\cite{Sels_2021} that this peak represents the smoking gun of many-body ergodicity breaking.
We then extend the study of the quantum sun model to the energy-driven ergodicity breaking transition, which emerges due to the existence of a many-body mobility edge~\cite{pawlik_sierant_2024}.
We show that the fidelity susceptibility exhibits a peak that to a high precision coincides with the position of the mobility edge.
The existence of mobility edges, both of single-particle~\cite{giuliano_delande_14, delande_pasek_17, bloch_luschen_18, sarma_li_20, bloch_luschen_18, sarma_li_20, gadway_an_2021, suotang_wang_2022, sarma_vu_2023, liu_wang_2020, liu_zhou_2023, liu_zhang_24, gao_khaymovich_25} and many-body origin~\cite{altshuler_levitov_97,kjall_pollman_2014,luitz_alet_2015,mondragon_laumann_2015,schiulaz_deroeck_16,enss_sirker_2017,villalonga_clark_2018,chanda_zakrzewski_2020,lin_2023,cadez_dietz_24,pawlik_sierant_2024}, have been extensively studied in the past, and the results of our work introduce a measure for detecting their structure in an interacting system.
We interpret our results using fading ergodicity, which captures the behavior of fidelity susceptibility as the system approaches the ergodicity-breaking critical point from the ergodic side, including the emergence of a maximally divergent peak associated with the onset of maximal chaos.

The paper is organized as follows.
In Sec.~\ref{sec:fading} we introduce the fading ergodicity scenario and the underlying quantum sun model, while in Sec.~\ref{sec:fidelity} we introduce the fidelity susceptibilities and formulate predictions for their scaling in the fading ergodicity regime.
We then numerically test these predictions in Sec.~\ref{sec:numerics}, focusing in Sec.~\ref{sec:ebt_interactions} on the interaction-driven ergodicity breaking transition in the middle of the spectrum, and in Sec.~\ref{sec:ebt_energy} on the energy-driven ergodicity breaking transition.
We conclude in Sec.~\ref{sec:conclusions}.

%%%%%%%%%%%%%%%%%%%%%%%%%%%%%%%%%%%%%%%%%%%%%%%%%%%%%%%%%%
\section{Breakdown of ergodicity} \label{sec2}

%%%%%%%%%%%%%%%%%%%%%%%%%%%%%%%%%%%%%%%%%%%%%%%%%%%%%%%%%%
\subsection{Fading ergodicity and quantum sun model} \label{sec:fading}

Fading ergodicity is a scenario for the breakdown of the conventional ETH from Eq.~\eqref{def_eth}.
In particular, it establishes a link between the conventional ETH and the complete breakdown of ETH via the gradual softening of fluctuations of diagonal and low-$\omega$ off-diagonal matrix elements~\cite{kliczkowski_vidmar2024}.
Since the latter are the focus of this work, we express the scaling of the typical matrix element, $V_{nm}\equiv \langle n|\hat V|m\rangle$, at $\bar E = (E_n+E_m)/2$ and $\omega = E_n-E_m \approx \omega_H$ as
\begin{equation}
\label{eq:matrix_elements}
    |V_{nm}|^2 \,(\omega\approx\omega_H,\bar E) \propto \rho(\bar E)^{-2/\eta} \;,
\end{equation}
where $\omega_H$ (setting $\hbar\equiv 1$) is the Heisenberg energy that corresponds to the mean level spacing.
The key quantity in Eq.~\eqref{eq:matrix_elements} is the fluctuation exponent $\eta$, which is a smooth function  that interpolates between $\eta = 2$ in the conventional ETH regime and $\eta \rightarrow \infty$ at the ergodicity breaking critical point.
Hence, $\eta = \infty$ at the critical point signals the complete breakdown of the ETH, i.e., the absence of exponential suppression of $|V_{nm}|^2$ with $L$.

In passing, we note that while Eq.~\eqref{eq:matrix_elements} describes deviations from the conventional ETH at any $\eta>2$, the behavior of the observable matrix elements at $\eta< \infty$ are still consistent with thermalization, and the short-range spectral statistics complies with predictions of random matrix theory ensembles~\cite{kliczkowski_vidmar2024}.
In this respect, the fading ergodicity regime is fundamentally different from non-ergodic regimes that exhibit different versions of {\it weak} ETH, such as quadratic fermionic models~\cite{lydzba_swietek_24} or integrable interacting models~\cite{biroli_lauchli_10, ikeda_ueda_15, alba_15, leblond_mallayya_19, brenes_leblond_20}.

The quantum sun model~\cite{suntajs_vidmar_22, suntajs_hopjan_24}, which was initially studied as the toy model of the avalanche theory~\cite{deroeck_huveneers_17}, represents a paradigmatic model of fading ergodicity~\cite{kliczkowski_vidmar2024}.
This model describes a system of spin-1/2 particles, with $N$ particles inside the thermal quantum dot coupled to $L$ particles outside the dot.
Its Hamiltonian can be written as~\cite{suntajs_hopjan_24}
\begin{equation}\label{eq:qsun}
    \hat{H} = \hat{R} + g_0\sum_{\ell=1}^{L}\alpha^{u_\ell}\hat{S}^x_{n(\ell)}\hat{S}^x_\ell + \sum_{\ell=1}^{L} h_\ell\hat{S}^z_\ell\;,
\end{equation}
where $\hat{R}$ describes the thermal quantum dot, i.e., it can be written as a $2^N\times 2^N$ random matrix drawn from the Gaussian orthogonal ensemble (GOE) in the computational basis so that it includes all-to-all interactions within the dot (we set $N=3$).
The particles outside the dot are subject to a random transverse field with $h_\ell$ being an independent and identically distributed (i.i.d.) random number in the interval $h_\ell\in[0.5,1.5]$.
The second term on the r.h.s.~of Eq.~(\ref{eq:qsun}) describes interactions between a particle outside the dot, with index $\ell$, and a randomly selected particle within the dot, with index $n(\ell$).
We set $g_0=1$.
The interactions decay exponentially with distance $u_\ell$, where $u_\ell$ is an i.i.d.~random number in the interval $u_\ell \in [(\ell-1)-0.2,(\ell-1)+0.2]$, except for $\ell=1$ when $u_1 = 0$.
The model is expected to exhibit the ergodicity-breaking phase transition in the thermodynamic limit at $\tilde{\alpha}_c=1/\sqrt{2}$~\cite{deroeck_huveneers_17, luitz_huveneers_17}.
At $\alpha\lesssim\alpha_c$, the system is nonergodic and it exhibits Fock-space localization, while at $\alpha\gtrsim\alpha_c$ it exhibits ergodic behavior~\cite{suntajs_hopjan_24}.
Throughout the work, we consider systems up to $L_{\rm tot}=L+N = 16$ spin-1/2 particles, which corresponds to the Hilbert-space dimension $D = 2^{L+N}=2^{16}$.

Exact numerical studies in finite systems~\cite{suntajs_hopjan_24, kliczkowski_vidmar2024, swietek_vidmar_scaling24} suggest the transition point, $\alpha_c$, to occur at a slightly larger value than predicted theoretically, $\tilde{\alpha}_c=1/\sqrt{2}$~\cite{deroeck_huveneers_17}.
In this paper, we take the numerically extracted critical point, $\alpha_c = 0.734$, which was obtained by performing a data collapse of the eigenstate entanglement entropy of a single site~\cite{swietek_vidmar_scaling24}.
That said, we have no particular arguments to expect $\alpha_c$ to differ from $\tilde{\alpha}_c$ in the thermodynamic limit, and the characterization of their differences is beyond the scope of this work.
Here, our focus is to compare the positions of the peak of fidelity susceptibility to $\alpha_c$, and to show consistency of different indicators to detect the critical point in finite systems.

%%%%%%%%%%%%%%%%%%%%%%%%%%%%%
\begin{figure}[t!]
\centering
\includegraphics[width=\columnwidth]{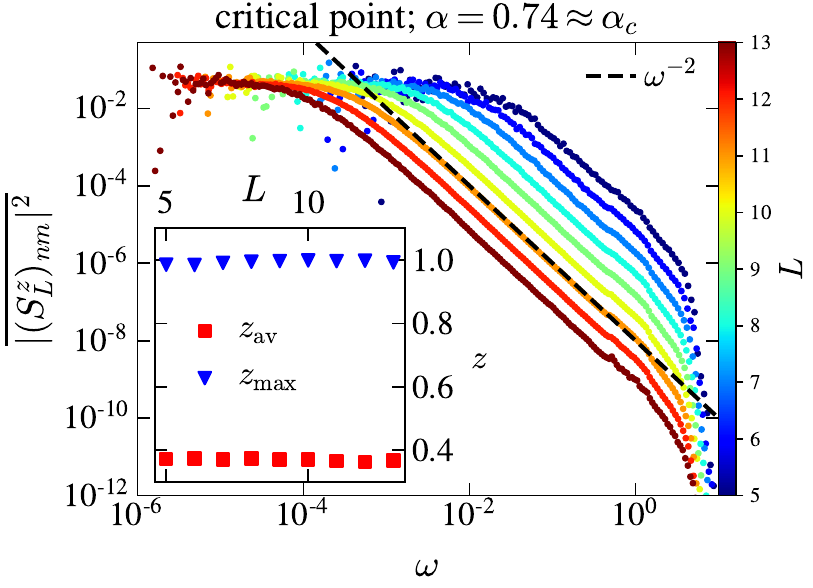}
\caption{
Matrix elements in the quantum sun model at $\alpha=0.74$, i.e., in the vicinity of the critical point ($\alpha_c = 0.734$).
Main panel:
Coarse-grained off-diagonal matrix elements $\overline{|(S_L^z)_{nm}|^2}$ vs $\omega$, at different system sizes $L$. 
We consider off-diagonal matrix elements with mean energies in the middle of the spectrum, and $100-2000$ realizations of the model, see Appendix~\ref{sec:technical} for details.
%Low-$\omega$ off-diagonal matrix elements form a plateau, with a width proportional to $\omega_H$ and a height independent of $L$.
Inset: Scaling of the eigenstate-to-eigenstate fluctuations of $500$ diagonal matrix elements in the middle of the spectrum, vs $L$.
The average value, $z^\text{av}$, is marked with red squares, while the maximal outlier, $z^\text{max}$, is marked with blue triangles. 
}
\label{fig1}
\end{figure}
%%%%%%%%%%%%%%%%%%%%%%%%%%%%%

A paradigmatic example of the observable that exhibits the scaling of matrix elements fluctuations according to Eq.~\eqref{eq:matrix_elements} is the operator $\hat S_{\ell=L}^z$ of the most distant particle from the dot.
While Ref.~\cite{kliczkowski_vidmar2024} systematically studied the matrix elements fluctuations in the fading ergodicity regime (i.e., at $\alpha>\alpha_c$), in the main panel of Fig.~\ref{fig1} we show the behavior of off-diagonal matrix elements $|(S_L^z)_{nm}|^2$ versus $\omega$ in the immediate vicinity of the critical point, at $\alpha=0.74$.
At $\omega\approx\omega_H$, the coarse-grained off-diagonal matrix elements exhibit a plateau, while at $\omega\gg \omega_H$ they exhibit the same polynomial decay as a Lorentzian function.
The height of the plateau at $\omega\approx\omega_H$ is independent of system size, which corresponds to the complete ETH breakdown and $\eta\to\infty$ in Eq.~\eqref{eq:matrix_elements}.
Another perspective of the ETH breakdown is shown in the inset of Fig.~\ref{fig1}, in which we show the average and maximal eigenstate-to-eigenstate fluctuations of diagonal matrix elements~\cite{kim_ikeda_14}.
We define $z_n=\ev{\hat S^z_L}{n+1}-\ev{\hat S^z_L}{n}$ to calculate
$z^\text{av}=\expval{z_n}_{n,\hat{H}}$ and $z^\text{max}=\expval{\text{max}(z_n)_n}_{\hat{H}}$, where $\expval{...}_{n,\hat{H}}$ refers to the averaging over a set of energy eigenstates and distinct Hamiltonian realizations, while $\expval{\text{max}(...)_n}_{\hat{H}}$ refers to selecting a maximal outlier over a set of energy eigenstates and then averaging over Hamiltonian realizations.
Results in the inset of Fig.~\ref{fig1} show that neither $z^\text{max}$ nor $z^\text{av}$ vanish in the thermodynamic limit $L\to\infty$.

%%%%%%%%%%%%%%%%%%%%%%%%%%%%%%%%%%%%%%%%%%%%%%%%%%%%%%%%%%
\subsection{Fidelity susceptibilities and AGP norms} \label{sec:fidelity}

We now turn our attention to small perturbations of the Hamiltonian, $\hat{H}(\lambda+\delta\lambda)=\hat{H}(\lambda)+\delta\lambda \hat{V}$, as discussed in the context of Eq.~\eqref{eq:fidelity}.
%as discussed in the context of Eq.~\eqref{eq:fidelity} in Introduction, Sec.~\ref{sec:introduction}.
The sensitivity of an arbitrary Hamiltonian eigenstate $|n(\lambda)\rangle$, with $\hat{H}(\lambda)\ket{n(\lambda)}=E_n(\lambda)\ket{n(\lambda)}$, to an infinitesimally small change of $\lambda$ can be quantified by the fidelity susceptibility~\cite{you_gu_2007},
\begin{equation}\label{eq:fidelity_excited}
    \chi_n=\sum_{m\neq n}\frac{|V_{nm}|^2}{\qty( E_m-E_n )^2}\;,
\end{equation}
which can be seen as a generalization of Eq.~\eqref{eq:fidelity} to excited states. From now on, we omit $\lambda$ in the notation of eigenstates and their eigenvalues.
In practice, since for some eigenstates $|n\rangle$ the nearest level spacings can be very small and may give rise to very large values of $\chi_n$, one usually calculates the typical fidelity susceptibility over Hamiltonian eigenstates,
\begin{equation}
\label{eq:fidelity:typical}
    \chi^{\rm typ}= e^{\expval{\ln{\chi_n}}_n} = e^{(1/D)\sum_{n=1}^D \ln{\chi_n}}\;.
\end{equation}

Another perspective is to interpret $\chi_n$ from Eq.~\eqref{eq:fidelity_excited} as 
$\chi_n = \sum_{m\neq n} |\mel{n}{\hat{\mathcal{A}}}{m}|^2$,
where $\hat{\mathcal{A}}$ stands for the AGP and is defined as $\hat{\mathcal{A}}|n\rangle=i\partial_\lambda|n\rangle$~\cite{kolodrubetz_sels_17}.
In particular, we consider its regularized version~\cite{sels_pandey_2020},
\begin{equation}\label{eq:fidelity:reg}
    \chi^r_n=\sum_{m\neq n}\frac{\omega_{nm}^2\abs{V_{nm}}^2}{\qty( \omega_{nm}^2 +\mu^2 )^2}=\sum_{m\neq n} \abs{\mel{n}{\hat{\mathcal{A}}^r}{m}}^2,
\end{equation}
where $\omega_{nm}=E_m-E_n$. We emphasize that the AGP at $\mu\neq 0$ can be expressed as
\begin{equation} \label{def_agp_reg}
    \hat{\mathcal{A}}^r=-\frac{1}{2}\int_{-\infty}^{\infty} dt\;\text{sgn}(t) e^{-\mu|t|} \hat{V}(t)\;,
\end{equation}
where $\hat{V}(t)$ is written in the Heisenberg picture, while $\mu$ is a characteristic energy scale and its inverse plays the role of a cutoff time. 
The choice of $\mu$ in numerical calculations will be discussed below.
The norm of AGP at $\mu\neq 0$ is defined as the average over Hamiltonian eigenstates,
\begin{equation}
\label{eq:fidelity:average}
    \chi^{\rm av}=\expval{\chi_n^r}_{n} = \frac{1}{D} \sum_{n=1}^D \chi_n^r = ||\hat{\mathcal{A}}^r||^2\;.
\end{equation}
We note that whenever we refer to $\chi^{\rm av}$, we have in mind its regularized version from Eq.~\eqref{eq:fidelity:average}. 
In all numerical calculations carried out in Sec.~\ref{sec:numerics}, we study $\chi^{\rm typ}$ and $\chi^{\rm av}$ as functions of the interaction $\alpha$ in the quantum sun model, see Eq.~\eqref{eq:qsun}.

To discuss the expected scaling in the fading ergodicity regime, we focus on $\chi^{\rm av}$ from Eq.~\eqref{eq:fidelity:average}.
To this end, we first convert sums in Eqs.~\eqref{eq:fidelity:reg} and~\eqref{eq:fidelity:average} into integrals and perform a variable substitution,
\begin{align} \label{def_chiav_1}
    \chi^{\rm av} = \frac{1}{D} \int dE \rho(E) & \int d\omega \rho(E+\omega) \\ \nonumber
    &\frac{\pi}{2\mu} \delta(\omega-\mu) |V(E+\omega/2,\omega)|^2 \;,
\end{align}
where we expressed the matrix element $|V_{nm}|^2$ by its coarse-grained value $|V(\bar E,\omega)|^2$, with $\bar E = E + \omega/2$, and we defined the delta function as $\delta(\omega-\mu) = (2/\pi)\mu \omega^2/(\omega^2+\mu^2)^2$. Additionally, $\rho(E)\propto D$ corresponds to the density of states at energy $E$.
The application of $\delta(\omega-\mu)$ for $\mu\ll 1$ simplifies Eq.~\eqref{def_chiav_1} to
\begin{equation}
    \chi^{\rm av} \approx \frac{\pi}{2\mu D} \int dE \rho(E) \rho(E+\mu) |V(E+\mu/2,\mu)|^2 \;.
\end{equation}
The latter expression is governed by the values at the peak of the density of states in the middle of the spectrum, at $E = E_0 = {\rm Tr}\{\hat H\}/D$.
Moreover, since we expect $\mu$ to be very small, we approximate $\rho(E) \approx \rho(E+\mu)$.
Then,
\begin{align}
    \chi^{\rm av}(E_0) & \propto \frac{1}{\mu D} \rho(E_0)^2 |V(E_0+\mu/2,\mu)|^2 \nonumber \\
    & \approx \frac{\rho(E_0)}{\mu} |V(E_0+\mu/2,\mu)|^2 \;, \label{def_chiav_3}
\end{align}
where we assumed $\rho(E_0) \approx D$, up to polynomial corrections in $L$.

At this point, one should apply the fading ergodicity ansatz for the fluctuations of the matrix elements from Eq.~\eqref{eq:matrix_elements}, in which $E_n = E_0$ and $E_m = E_0 + \mu$.
This step, however, also requires certain insight about the optimal value of $\mu$.
Deep in the ergodic regime, the value of $\mu$ is usually set such that it lies within the low-$\omega$ plateau of the coarse-grained off-diagonal matrix elements~\cite{sels_pandey_2020}.
In the fading ergodicity regime, the width of the low-$\omega$ plateau is roughly given by the Thouless energy $\Gamma$~\cite{kliczkowski_vidmar2024}, and hence $\mu$ may in principle take any value in the interval $\omega_H \lesssim \mu \ll \Gamma$.
Nevertheless, in the vicinity of the critical point, the Thouless energy shrinks to the Heisenberg energy, $\Gamma\to \omega_H$~\cite{suntajs_vidmar_22}, which manifests itself as the low-$\omega$ plateau region being very narrow, see the main panel of Fig.~\ref{fig1}.
This narrows the choice of $\mu$ to the vicinity of $\omega_H$, up to polynomial corrections in $L$.
Therefore, we set $\mu \propto \omega_H$ in Sec.~\ref{sec:numerics}, while we also test possible polynomial corrections to this scaling in Appendix~\ref{sec:technical}.

Applying the fading ergodicity ansatz from Eq.~\eqref{eq:matrix_elements} and setting $\mu = \omega_H$, we can convert Eq.~\eqref{def_chiav_3} to a prediction for the scaling of the AGP norm in the fading ergodicity regime, $\chi^{\rm av} \to \chi_{\rm fading}$, with
\begin{equation} \label{def_chi_fading}
     \chi_{\rm fading}(E_0) \propto \frac{\rho(E_0)^{1-2/\eta}}{\mu} = \chi_{\rm ETH}\, \rho(E_0)^{1-2/\eta}\;,
\end{equation}
where we defined $\chi_{\rm ETH} = 1/\mu$.
Furthermore, since $\mu = \omega_H \propto \rho(E_0)^{-1}$, we arrive at
\begin{equation} \label{def_chi_fading_2}
     \chi_{\rm fading}(E_0) \propto \rho(E_0)^{2-2/\eta} \;,
\end{equation}
which is the main result of this section and will be tested numerically in Sec.~\ref{sec:ebt_interactions}.
It formulates the closed-from expression for $\chi^{\rm av}$ in the entire fading ergodicity regime, i.e., from the conventional ETH limit at $\alpha=1$ when $\eta=2$ and $\chi_{\rm fading}(E_0) \to \rho(E_0) \propto D$, to the ergodicity-breaking critical point at $\alpha=\alpha_c$ when $\eta\to\infty$, and hence $\chi_{\rm fading}(E_0) \to \rho(E_0)^2 \propto D^2$ saturates the upper bound.
We refer to the ergodicity-breaking critical point at which $\chi_{\rm fading}$ is maximal as the point of maximal quantum chaos.

While the above arguments have been built on properties of the average, $\chi^{\rm av}$, we expect similar scaling of both $\chi^{\rm av}$ and $\chi^{\rm typ}$ in the fading ergodicity regime.
Hence, we express the expected scaling of the typical fidelity susceptibility as
\begin{equation}
\label{eq:susceptibilities_fading2}
    \chi^\text{typ}\propto\omega_H^{-2+{2}/{\eta}}\propto \chi_{\rm fading}(E_0)\;,
\end{equation}
which can also be understood as the value governed by the nearest level spacing $\omega_H$ in Eq.~\eqref{eq:fidelity_excited}.

So far, we have considered the averages over all eigenstates, which effectively probe the properties of eigenstates in the middle of the spectrum where the density of states is maximal.
We then extend our analysis to microcanonical windows centered at energy $E$ away from the middle of the spectrum.
To this end, we generalize the typical fidelity susceptibility to
\begin{equation}\label{eq:fidelity:finiteE:typ}
    \chi^{\rm typ}(E)=\exp({\frac{1}{\mathcal{N}_{E,\Delta_\epsilon}}\sum_{\abs{E_n-E}<\Delta_\epsilon}\ln{\chi_n}})\;,
\end{equation}
where the average is performed within the microcanonical window at the target energy $E$ of width $2\Delta_\epsilon$ and cardinality $\mathcal{N}_{E, \Delta\epsilon}$. 
In the numerical calculations, we introduce the scaled energy density for a given Hamiltonian realization,
\begin{equation} \label{def_energy_density}
\epsilon=(E-E_\text{min}) / (E_{\rm max}-E_{\rm min})\;, 
\end{equation}
where $E_{\rm min}$ and $E_{\rm max}$ are the lowest and highest eigenenergies of the Hamiltonian, respectively.
Consequently, $\chi^{\rm typ}(E) \to \chi^{\rm typ}(\epsilon)$ in Eq.~\eqref{eq:fidelity:finiteE:typ}, and we then further average $\chi^{\rm typ}(\epsilon)$ over Hamiltonian realizations.
However, to predict the scaling of $\chi^{\rm typ}(\epsilon)$ with system size $L$ and interaction $\alpha$, one needs to understand the dependence of the fluctuation exponent $\eta$ on $\epsilon$.
This behavior has not yet been studied before, and we focus on this question in Sec.~\ref{sec:ebt_energy}.

%%%%%%%%%%%%%%%%%%%%%%%%%%%%%%%%%%%%%%%%%%%%%%%%%%%%%%%%%%
\section{Numerical results} \label{sec:numerics}

\subsection{Interaction-driven ergodicity breaking} \label{sec:ebt_interactions}

We first study the ergodicity-breaking transition in the middle of the spectrum, at $E=E_0$.
We focus on the perturbation $\hat{V}=\hat{S}_{L}^{z}$, i.e., on the $z-$projection of the spin that resides at the largest distance from the quantum dot, cf.~$\ell=L$ in Eq.~\eqref{eq:qsun}.
We will later complement our results by studying other perturbations.
We set the cutoff energy $\mu=\sqrt{L_{\rm tot}}/D$ where $L_{\rm tot}=L+N$, such that it is proportional to the mean level spacing and the inverse density of states in the middle of the spectrum, $\mu\propto \omega_H \propto \rho(E_0)^{-1}$, see also Appendix~\ref{sec:technical}.
%(up to polynomial corrections). 
In contrast to Eqs.~\eqref{eq:fidelity:typical} and~\eqref{eq:fidelity:average} where the averages of $\chi^{\rm typ}$ and $\chi^{\rm av}$ are taken over all Hamiltonian eigenstates, we here minimize finite-size effects by only averaging over $50\%$ of eigenstates in the middle of the spectrum (we refer to $\chi^{\rm typ}$ and $\chi^{\rm av}$ as the typical and average fidelity susceptibility).
We then further average the results over $10^2-10^4$ Hamiltonian realizations, see Appendix~\ref{sec:technical} for further details. 

%%%%%%%%%%%%%%%%%%%%%%%%%%%%%
\begin{figure}[t!]
\centering
\includegraphics[width=\columnwidth]{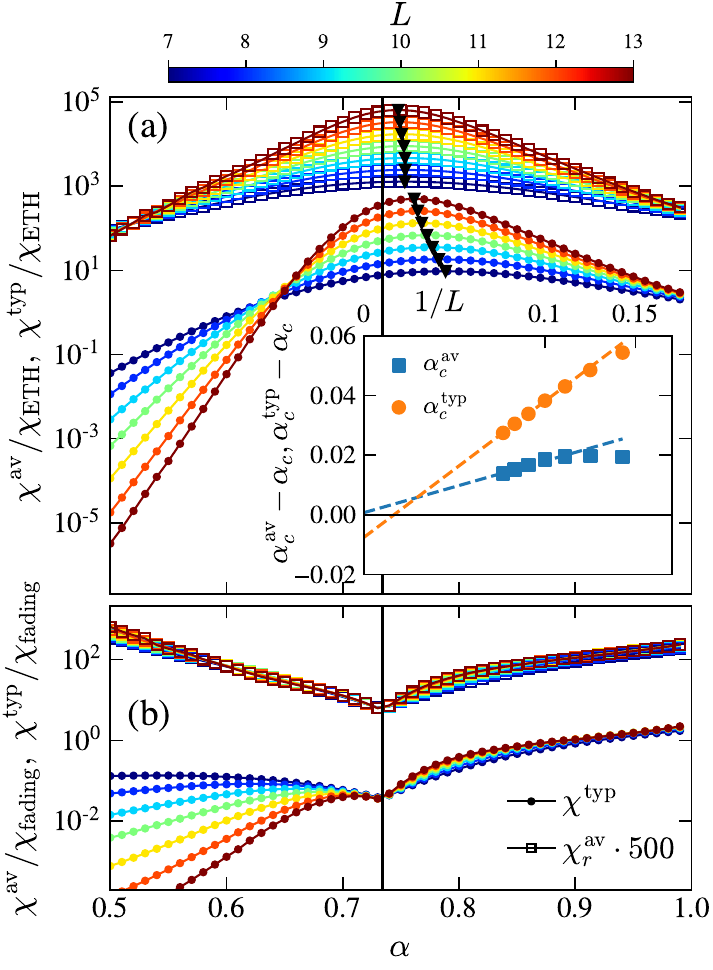}
\caption{
(a)~Fidelity susceptibilities $\chi^\text{typ}/\chi_\text{ETH}$ and $\chi^\text{av}/\chi_\text{ETH}$ as functions of the interaction strength $\alpha$. Black triangles mark the positions of the peak, $\alpha_c^\text{typ}$ and $\alpha_c^\text{av}$, established from the fifth-order polynomial fits to numerical results, and the vertical solid line is in both panels located at $\alpha = \alpha_c = 0.734$~\cite{swietek_vidmar_scaling24}.
The inset shows the scaling of $\alpha_c^\text{typ}-\alpha_c$ and $\alpha_c^\text{av}-\alpha_c$ versus $1/L$, and dashed lines are linear extrapolations from $L\ge 9$. 
(b)~Scaled fidelity susceptibilities $\chi^\text{typ}/\chi_\text{fading}$ and $\chi^\text{av}/\chi_\text{fading}$ as functions of the interaction strength $\alpha$, where $\chi_{\rm fading}$ is given by Eq.~\eqref{def_chi_fading_2} upon replacing the proportionality with equality, and $\eta$ given by Eq.~\eqref{eq:eta} on the ergodic side, while on the nonergodic side we take $\eta\to-\eta$.
}
\label{fig2}
\end{figure}
%%%%%%%%%%%%%%%%%%%%%%%%%%%%%

Figure~\ref{fig2}(a) shows numerical results for the scaled fidelity susceptibilities $\chi^\text{typ}/\chi_\text{ETH}$ and $\chi^\text{av}/\chi_\text{ETH}$, for the operator $\hat S_L^z$, versus the interaction strength $\alpha$. 
Remarkably, both quantities exhibit peaks very close to the critical value $\alpha_c=0.734$, which is shown as a vertical solid line.
For a given system size $L$, we determine the positions of peaks of $\chi^\text{typ}/\chi_\text{ETH}$ and $\chi^\text{av}/\chi_\text{ETH}$, denoted as $\alpha_c^\text{typ}$ and $\alpha_c^\text{av}$, respectively, from the fifth-order polynomial fits to numerical results.
The corresponding values of $\chi^\text{typ}/\chi_\text{ETH}$ and $\chi^\text{av}/\chi_\text{ETH}$ at the peak are marked by black triangles in Fig.~\ref{fig2}(a).
The finite-size scalings of the differences $\alpha_c^\text{typ}-\alpha_c$ and $\alpha_c^\text{av}-\alpha_c$ are shown in the inset of Fig.~\ref{fig2}(a).
The linear extrapolation to the thermodynamic limit, $1/L \to 0$, suggest the differences to completely vanish or to become vanishingly small. 
This result represents evidence that the position of the peak of fidelity susceptibilities likely coincides with the ergodicity-breaking critical point in the thermodynamic limit.

To test the scaling of fidelity susceptibilities in the fading ergodicity regime, predicted by Eqs.~\eqref{def_chi_fading} and~\eqref{eq:susceptibilities_fading2}, one needs the knowledge of the function $\eta(\alpha)$ from Eq.~(\ref{eq:matrix_elements}).
The prediction for $\eta(\alpha)$ was already formulated in Ref.~\cite{kliczkowski_vidmar2024}.
It builds on the surmise for the scaling of the low-$\omega$ off-diagonal matrix elements, 
$|V_{nm}|^2 = \omega_H/\Gamma$, where $\Gamma$ is the Thouless energy~\cite{kliczkowski_vidmar2024}.
Then, expressing $\omega_H \propto 2^{-L} = \exp\{-L \ln(1/\tilde\alpha_c^2)\} $, where $\tilde\alpha_c=1/\sqrt{2}$, and $\Gamma \propto \exp\{-L\ln(1/\alpha^2)\}$~\cite{suntajs_vidmar_22}, one obtains $|V_{nm}|^2 \propto \exp\{-L \ln(\alpha^2/\tilde\alpha_c^2)\}$.
Inserting the latter into the fading ergodicity ansatz from Eq.~\eqref{eq:matrix_elements},
one obtains
\begin{equation} \label{eq:eta}
    \eta(\alpha) = 2\left(1-\frac{\ln\alpha}{\ln\alpha_c}\right)^{-1}\;,
\end{equation}
in which we have replaced the analytically predicted value for the critical point, $\tilde{\alpha}_c$, with the numerical obtained value, $\alpha_c$.

In Fig.~\ref{fig2}(b) we plot $\chi^\text{typ}/\chi_\text{fading}$ and $\chi^\text{av}/\chi_\text{fading}$ versus $\alpha$, where $\chi_{\rm fading}$ was introduced in Eq.~\eqref{def_chi_fading_2} and $\eta(\alpha)$ is given by Eq.~\eqref{eq:eta}.
We find a reasonably good data collapse for different system sizes $L$ in the entire fading ergodicity regime, $\alpha > \alpha_c$, confirming that indeed the prediction for $\chi_{\rm fading}$ in Eq.~\eqref{def_chi_fading_2}, despite being rather simple, still accurately describes the numerical results.

Extending our analysis to the nonergodic side, we observe that $\chi^\text{av}/\chi_\text{fading}$ still exhibits a good data collapse for different $L$ in a rather broad regime of $\alpha < \alpha_c$.
This scale-invariant behavior of $\chi^\text{av}/\chi_\text{fading}$ may appear surprising provided that no theoretical framework to describe the fluctuations of matrix elements on the nonergodic side has so far been established.
Associating $\eta$ with the correlation/localization length $\xi$, this suggest, as already observed in~\cite{suntajs_vidmar_22, swietek_vidmar_scaling24}, that the divergence of $\xi$ is identical on both sides of the transition.
However, for the typical fidelity susceptibility, $\chi^\text{typ}/\chi_\text{fading}$ does not exhibit any scale-invariant behavior at $\alpha < \alpha_c$ in Fig.~\ref{fig2}(b).
The different behavior between $\chi^\text{av}$ and $\chi^\text{typ}$ on the nonergodic side is intriguing and deserves more attention in future work.

%%%%%%%%%%%%%%%%%%%%%%%%%%%%%
\begin{figure}[t!]
\centering
\includegraphics[width=\columnwidth]{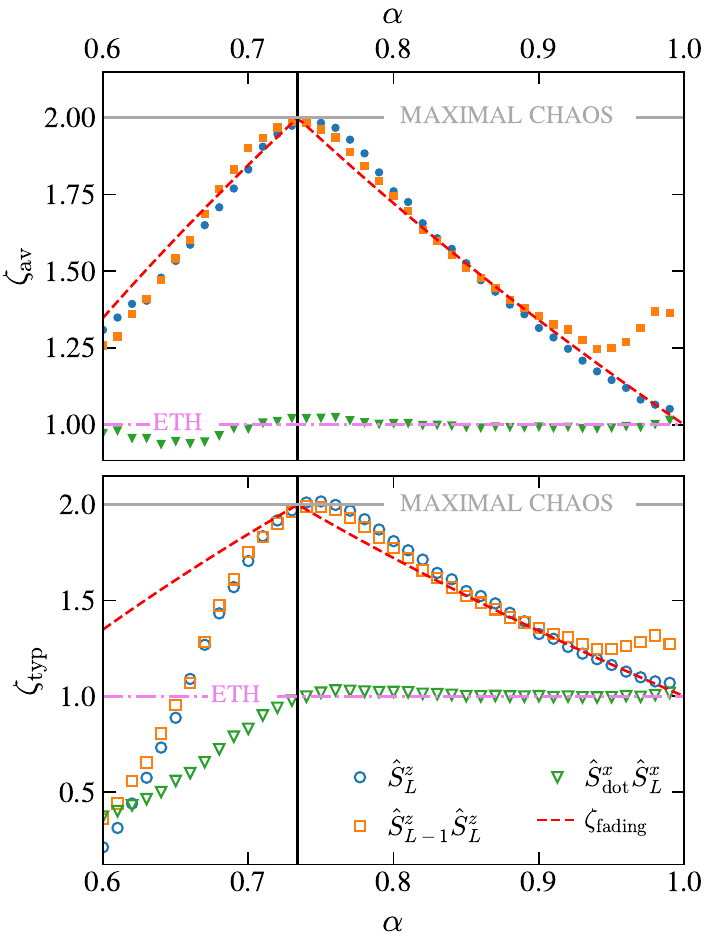}
\caption{Exponents (a) $\zeta_\text{av}$ and (b) $\zeta_\text{typ}$, see Eq.~\eqref{def_zeta}, versus the interaction strength $\alpha$, for three different observables.
The dashed-dotted horizontal line marks the ETH prediction, $\zeta_{\rm av/typ}=1$, while the solid horizontal line marks the expectation for the maximal chaos, $\zeta_{\rm av/typ}=2$.
The dashed lines indicate the exponent established from fading ergodicity,
$\zeta_{\rm fading}=2-2/\eta=1+\ln{\alpha}/\ln{\alpha_c}$ for $\alpha>\alpha_c$ and $\zeta_{\rm fading}=3-\ln{\alpha}/\ln{\alpha_c}$ for $\alpha<\alpha_c$, see Eq.~\eqref{eq:eta}.
The solid vertical line indicates the critical point $\alpha_c=0.734$~\cite{swietek_vidmar_scaling24}.
}
\label{fig5}
\end{figure}
%%%%%%%%%%%%%%%%%%%%%%%%%%%%%

So far, we have considered the system's response to the perturbation described by $\hat S_L^z$.
The motivation for choosing the latter operator is its strong sensitivity to the ergodicity-breaking transition~\cite{suntajs_hopjan_24}.
Therefore, the behavior of the corresponding fidelity susceptibilities on the nonergodic side is similar (but it is not equal) to those studied in integrable systems, in particular, to those that are characterized as integrability-preserving~\cite{sels_pandey_2020} and, eventually, weak integrability-breaking perturbations~\cite{orlov_tiutiakina_23, pozsgay_sharipov_24, PhysRevLett.96.067202, kurlov_malikis_22, surace_motrunich_23, vanovac_surace_24}.
Here we complement our study by considering two other perturbations, described by $\hat{S}^z_{L-1}\hat{S}^z_{L}$ and $\hat{S}^x_{\rm dot}\hat{S}^x_L$.
The study of the matrix elements of these observables~\cite{kliczkowski_vidmar2024} showed that the former exhibits a scaling similar to the one of $\hat S_L^z$, i.e., it exhibits features of fading ergodicity, while the latter does not.
In particular, the matrix element fluctuations of $\hat{S}^x_{\rm dot}\hat{S}^x_L$ are consistent with the conventional ETH in the entire ergodic phase as well as on the nonergodic side~\cite{kliczkowski_vidmar2024}.
Hence, this observable is insensitive to the ergodicity-breaking transition, and we expect the corresponding fidelity susceptibilities to satisfy the scaling predicted by $\chi_{\rm ETH}$ for a broad range of $\alpha$.
One may also say that $\hat{S}^x_{\rm dot}\hat{S}^x_L$ shares similarities with the behavior of strong integrability-breaking perturbations.

In the numerical calculations, we pursue the following approach.
At a given interaction $\alpha$, we vary the system size $L$ and determine the exponent $\zeta$ from the ansatz
\begin{equation} \label{def_zeta}
\chi^{\rm av/typ}(E_0) \propto \rho(E_0)^{\zeta_{\rm av/typ}}\;.
\end{equation}
If the fading ergodicity prediction applies, cf.~Eq.~\eqref{def_chi_fading_2}, then $\zeta = 2-2/\eta$.
Hence, $\zeta = 1$ marks the conventional ETH and $\zeta = 2$ indicates the emergence of maximal chaos at the ergodicity-breaking critical point.

We show $\zeta_{\rm av}$ and $\zeta_{\rm typ}$ as functions of $\alpha$ in Figs.~\ref{fig5}(a) and~\ref{fig5}(b), respectively.
As expected from the discussion above, for the observables $\hat{S}^z_{L}$ and $\hat{S}^z_{L-1}\hat{S}^z_{L}$ they smoothly transition from $\zeta_{\rm av/typ}\approx 1$ deep in the ergodic phase~\footnote{At $\alpha=1$ we note small deviations from the ETH observed for certain observables, see also Ref.~\cite{swietek_vidmar_scaling24}, which remains an open question for further research.} to $\zeta_{\rm av/typ} = 2$ at the ergodicity-breaking critical point.
The dashed lines in Fig.~\ref{fig5} denote the fading ergodicity prediction $\zeta_{\rm av/typ} = 2-2/\eta$, which provides an accurate description of the numerical results for $\hat{S}^z_{L}$ and $\hat{S}^z_{L-1}\hat{S}^z_{L}$ in the entire ergodic phase, and for $\zeta_{\rm av}$ even on the nonergodic side.
On the other hand, $\zeta_{\rm av} \approx \zeta_{\rm typ} \approx 1$ across the entire ergodic phase for $\hat{S}^x_{\rm dot}\hat{S}^x_L$, and $\zeta_{\rm av} \approx 1$ also on the nonergodic side. 
This confirms the expectations that the later operator is not a good indicator of the ergodicity-breaking transition.

%%%%%%%%%%%%%%%%%%%%%%%%%%%%%
\begin{figure}[b]
\centering
\includegraphics[width=\columnwidth]{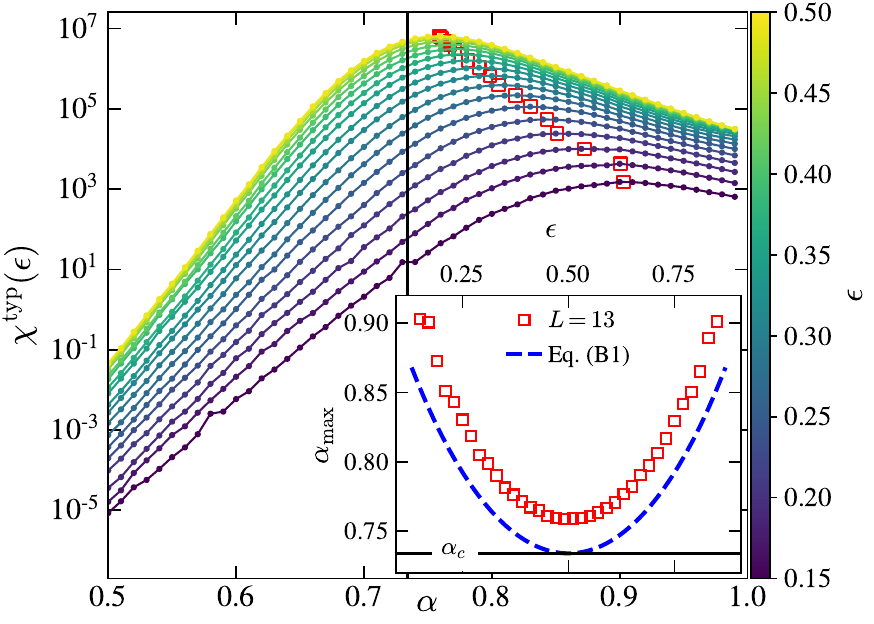}
\caption{
Main panel: typical fidelity susceptibility $\chi^\text{typ}(\epsilon)$ at energy density $\epsilon$ versus interaction $\alpha$, at system size $L=13$.
Brighter colors correspond to larger $\epsilon$. 
Inset: positions $\alpha_{\rm max}$ of peaks of $\chi^\text{typ}(\epsilon)$ extracted from the fifth-order polynomial fits to numerical results from the main panel.
The dashed line marks the analytical prediction for the many-body mobility edge in the thermodynamic limit, see Eq.~(\ref{eq:sm:mobility_edge}) in Appendix~\ref{sec:mobility_edge} and Ref.~\cite{pawlik_sierant_2024}.
The vertical solid line in the main panel and the horizontal solid line in the inset mark the numerical prediction of the critical point in the middle of the spectrum, $\alpha_c = 0.734$~\cite{swietek_vidmar_scaling24}.}
\label{fig3}
\end{figure}
%%%%%%%%%%%%%%%%%%%%%%%%%%%%%

%%%%%%%%%%%%%%%%%%%%%%%%%%%%%%%%%%%%%%%%%%%%%%%%%%%%%%%%%%
\subsection{Energy-driven ergodicity breaking} \label{sec:ebt_energy}

We now generalize our study to energies away from the middle of the spectrum.
Specifically, we consider the typical fidelity susceptibilities $\chi^{\rm typ}$ at energy $E$, for which the corresponding energy density $\epsilon$, see Eq.~\eqref{def_energy_density}, is below (or above) $\epsilon=0.5$.
We calculate $\chi^{\rm typ}(\epsilon)$ in a microcanonical energy window as defined in Eq.~\eqref{eq:fidelity:finiteE:typ}.
We also generalize the calculation of the level spacing to the corresponding energy density, $\omega_H \to \omega_H(\epsilon)$, and similarly for the density of states, $\rho \to \rho(\epsilon) = 1/\omega_H(\epsilon)$, see also Appendix~\ref{sec:technical} for details.

The main question that we address is whether the position of the peak of $\chi^{\rm typ}(\epsilon)$ shifts as a function of energy.
Recently, the shape of the many-body mobility edge has been established in the quantum sun model~\cite{pawlik_sierant_2024}.
It is then natural to ask whether the peak of $\chi^{\rm typ}(\epsilon)$ follows the shape of the many-body mobility edge.
Moreover, we also study how does the value of the peak of $\chi^{\rm typ}(\epsilon)$ scale with the system size, and to what extent one may define the notion of maximal chaos away from the middle of the spectrum.

%%%%%%%%%%%%%%%%%%%%%%%%%%%%%
\begin{figure}[t!]
\centering
\includegraphics[width=\columnwidth]{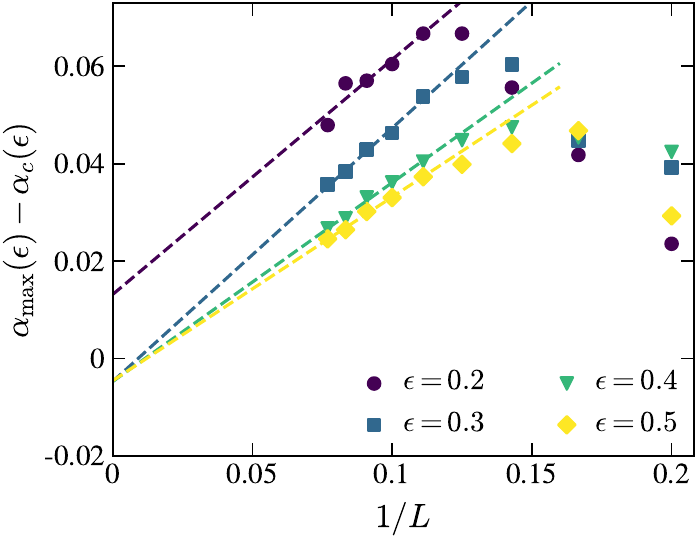}
\caption{
The difference $\alpha_{\rm max}(\epsilon) - \alpha_c(\epsilon)$ between the peak position of fidelity susceptibility, $\alpha_{\rm max}(\epsilon)$, and the position of the mobility edge, $\alpha_c(\epsilon)$ from Eq.~\eqref{eq:sm:mobility_edge}, plotted against $1/L$ at different energy densities $\epsilon$.
Dashed lines correspond to least-squares fits of $c_1+c_2/L$, where $c_1$ and $c_2$ are independent fitting parameters, to the numerical results with $L\ge 9$.
}
\label{figS3}
\end{figure}
%%%%%%%%%%%%%%%%%%%%%%%%%%%%%

The main panel of Fig.~\ref{fig3} shows $\chi^{\rm typ}(\epsilon)$ versus $\alpha$, for various $\epsilon \leq 0.5$ and a fixed system size $L=13$.
We observe the drift of the peak of $\chi^{\rm typ}(\epsilon)$, which emerges at $\alpha\approx\alpha_c$ for $\epsilon=0.5$, as discussed in Fig.~\ref{fig2}, to larger values of $\alpha$ with decreasing $\epsilon$.
We determine the peak position, $\alpha_{\rm max}(\epsilon)$, and the value of $\chi^{\rm typ}(\epsilon)$ at the peak, $\chi^{\rm typ}_{\rm max}(\epsilon)$, using a fifth-order polynomial fit to numerical data.
The values of $\chi^{\rm typ}_{\rm max}(\epsilon)$ are marked with red squares in the main panel of Fig.~\ref{fig3}.

Symbols in the inset of Fig.~\ref{fig3}(b) show $\alpha_{\rm max}$ versus $\epsilon$ at $L=13$.
Results are compared to the shape of the many-body mobility edge $\alpha_c(\epsilon)$ introduced in Ref.~\cite{pawlik_sierant_2024}, see also Eq.~(\ref{eq:sm:mobility_edge}) in Appendix~\ref{sec:mobility_edge}.
Both $\alpha_{\rm max}(\epsilon)$ and $\alpha_c(\epsilon)$ exhibit a very similar functional dependence on $\epsilon$.
However, while $\alpha_c(\epsilon)$ corresponds to the extrapolated result in the thermodynamic limit, $\alpha_{\rm max}(\epsilon)$ is a result for a finite system with $L=13$.
We therefore carry out the finite-size analysis of the difference $\alpha_{\rm max}(\epsilon) - \alpha_c(\epsilon)$, which is shown in Fig.~\ref{figS3}.
We fit the results using a linear function in $1/L$ and show a convincing evidence that $\alpha_{\rm max}(\epsilon) \approx \alpha_c(\epsilon)$ in the thermodynamic limit, with similar accuracy as for the results in the middle of the spectrum shown in the inset of Fig.~\ref{fig2}(a).
Small differences can only be observed at the lowest $\epsilon$ studied, $\epsilon = 0.20$, which we expect are finite-size effects due to a small number of available states in the corresponding microcanonical window.
These results suggest that, indeed, the peak of the fidelity susceptibility, $\chi^{\rm typ}(\epsilon)$, likely emerges at the many-body mobility edge throughout the spectrum.

%%%%%%%%%%%%%%%%%%%%%%%%%%%%%
\begin{figure}[t!]
\centering
\includegraphics[width=\columnwidth]{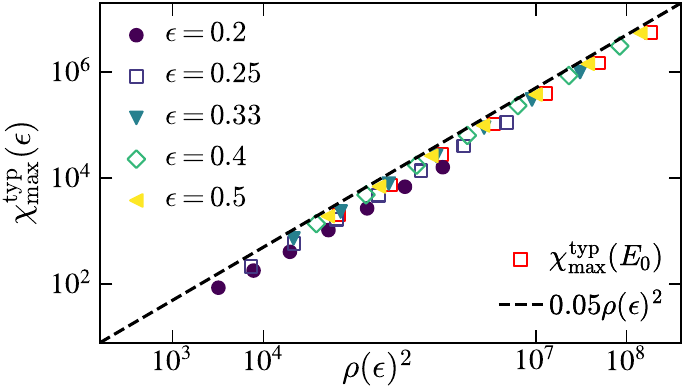}
\caption{
The scaling of the maximal values of fidelity susceptibilities, $\chi^\text{typ}_{\rm max}(\epsilon)$, with the square of density of states, $\rho(\epsilon)^2$, at different energy densities $\epsilon$. The scaling of $\chi_{\rm max}^{\rm typ}(E_0)$, calculated in the middle of the spectrum and presented in Fig.~\ref{fig2}, is included for comparison.
The solid line represents a linear function, $0.05\rho(\epsilon)^2$, and serves as a guide for the eye.
}
\label{fig4}
\end{figure}
%%%%%%%%%%%%%%%%%%%%%%%%%%%%%

Next we study how the peak height $\chi^{\rm typ}_{\rm max}(\epsilon)$ scales with system size $L$ at different energy densities $\epsilon$.
The results presented in the main panel of Fig.~\ref{fig3} indicate that $\chi^{\rm typ}_{\rm max}(\epsilon)$, at a fixed $L=13$, decreases when $\epsilon$ departs from the middle of the spectrum. This suggests that $\chi^{\rm typ}_{\rm max}(\epsilon)$ may scale differently with $L$ at different $\epsilon$.

In Fig.~\ref{fig4} we show results for $\chi^{\rm typ}_{\rm max}(\epsilon)$ at different system sizes $L$ and energy densities $\epsilon$.
We observe a remarkably good scaling collapse when $\chi^{\rm typ}_{\rm max}(\epsilon)$ is plotted as a function of $\rho(\epsilon)^2$, where $\rho(\epsilon)$ corresponds to the density of states at the given $\epsilon$.
In fact, $\chi^{\rm typ}_{\rm max}(\epsilon)$ increases linearly with $\rho(\epsilon)^2$ for all values of $\epsilon$ under considerations.
Remarkably, the results in Fig.~\ref{fig4} suggest a generalization of Eq.~\eqref{def_zeta} to energies away from the middle of the spectrum,
\begin{equation} \label{def_zeta_epsilon}
    \chi_{\rm max}^{\rm typ}(\epsilon) \propto \rho(\epsilon)^{\zeta_{\rm max}^{\rm typ}}\;,\;\;\;
    {\zeta_{\rm max}^{\rm typ}} = 2\;.
\end{equation}
Equation~\eqref{def_zeta_epsilon} can be interpreted as the notion of "maximal" chaos within an energy window away from the middle of the spectrum.

%%%%%%%%%%%%%%%%%%%%%%%%%%%%%
\begin{figure}[t!]
\centering
\includegraphics[width=\columnwidth]{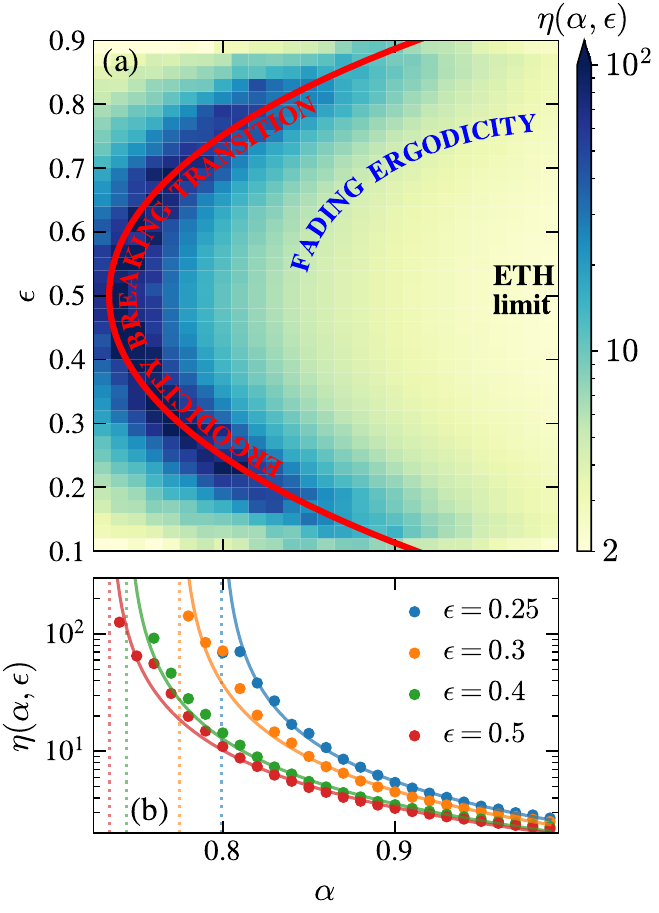}
\caption{
(a) Fluctuation exponent $\eta$ versus the interaction strength $\alpha$ and energy density $\epsilon$.
The ETH limit corresponds to $\eta=2$, fading ergodicity corresponds to $2<\eta<\infty$, while $\eta\to\infty$ denotes the ergodicity-breaking critical point.
Solid red line is the analytical prediction for the many-body mobility edge, see Eq.~\eqref{eq:sm:mobility_edge} in Appendix~\ref{sec:mobility_edge}. (b) Fluctuation exponent $\eta$ at fixed values of energy density $\epsilon$ as function of interaction $\alpha$. The solid lines represent fits of $A[\eta(\alpha,\epsilon)]^\nu$ to the results at $\eta\lesssim25$, with $\eta(\alpha,\epsilon)$ given by Eq.~\eqref{eq:eta:finiteE}. We find $A\in(0.95,1.25)$ and $\nu\approx1.3$. The vertical dotted lines show the prediction of Eq.~\eqref{eq:sm:mobility_edge}.
}
\label{figS5}
\end{figure}
%%%%%%%%%%%%%%%%%%%%%%%%%%%%%

Finally, we comment on the consequences of Eq.~\eqref{def_zeta_epsilon} for the properties of matrix elements of observables away from the middle of the spectrum. It suggests a complete breakdown of the ETH at the many-body mobility edge.
In the middle of the spectrum (at $\epsilon=0.5$), this corresponds to $\eta(\alpha\to\alpha_c) \to \infty$, see Eq.~\eqref{eq:eta}.
Here we generalize the function $\eta$ to energy densities $\epsilon\neq 0.5$, $\eta(\alpha)\rightarrow\eta(\alpha,\epsilon)$. We generalize the ansatz for the scaling of matrix elements from Eq.~\eqref{eq:matrix_elements} to
\begin{equation} \label{def_Vnm2_eta}
    \overline{|V_{nm}|^2}(\omega\approx\omega_H,\epsilon)\approx\frac{\omega_H(\epsilon)}{\Gamma(\epsilon)} \propto \rho(\epsilon)^{-2/\eta(\alpha,\epsilon)}\;,
\end{equation}
where $\overline{|V_{nm}|^2}$ corresponds to a coarse-grained matrix element, see Eq.~\eqref{eq:sm:eta:inf} in Appendix~\ref{sec:softening}, and $\omega_H(\epsilon)$ and $\Gamma(\epsilon)$ correspond to the Heisenberg and Thouless energy, respectively, at energy density $\epsilon$.

We first extract $\eta(\alpha,\epsilon)$ numerically via Eq.~\eqref{def_Vnm2_eta}, as discussed in details in Appendix~\ref{sec:softening}. 
Results are shown as a density plot in Fig.~\ref{figS5}(a).
Remarkably, the region in which $\eta$ diverges, see the dark blue region in Fig.~\ref{figS5}(a), very accurately agrees with the position of the many-body mobility edge from Eq.~\eqref{eq:sm:mobility_edge}, see the solid red line in Fig.~\ref{figS5}(a). 

The analytical calculation of $\eta(\alpha,\epsilon)$ requires the knowledge of both $\omega_H(\epsilon)$ and $\Gamma(\epsilon)$.
Here, we follow the approximation in which the energy dependence is taken into account in $\omega_H$, but not in $\Gamma$.
Specifically, since $\omega_H(\epsilon) \propto \rho(\epsilon)^{-1}$, we follow the hybridization condition in Ref.~\cite{pawlik_sierant_2024}, which in leading order of $1/L$  yields $\rho(\epsilon)^{-1}\propto\alpha_c(\epsilon)^{2L} \propto \exp{2L\ln \alpha_c(\epsilon)}$, where $\alpha_c(\epsilon)$ is given by Eq.~\eqref{eq:sm:mobility_edge}. 
We then rewrite the Thouless energy, introduced in the context of Eq.~\eqref{eq:eta}, as
$\Gamma\propto \exp{2L\ln(\alpha)} \propto \rho(\epsilon)^{-\ln\alpha/\ln\alpha_c(\epsilon)}$, and we obtain
\begin{equation}\label{eq:eta:finiteE}
    \eta(\alpha,\epsilon)=2\qty(1-\frac{\ln{\alpha}}{\ln{\alpha_c(\epsilon)}})^{-1}.
\end{equation}
This expression, while being rather simple, still contains the divergence of $\eta$ at the mobility edge at any finite energy density.
To describe the actual numerical results, however, we find in Fig.~\ref{figS5}(b) that a heuristic rescaling $\eta(\alpha,\epsilon)\to \eta(\alpha,\epsilon)^\nu$ provides a more accurate description.

The observations from Fig.~\ref{figS5} contain some important messages.
They establish the notion of fading ergodicity away from the middle of the spectrum, and hence generalize the results from Ref.~\cite{kliczkowski_vidmar2024}.
Moreover, they show that the peak of the fidelity susceptibility, and hence the ergodicity-breaking critical point, can be associated with the complete breakdown of ETH at arbitrary non-zero energy density.
This suggests a common theoretical framework for describing ergodicity-breaking transitions in the middle of the spectrum and in the low-energy part of the spectrum, and it calls for establishing a firmer connection between the ergodicity-breaking transitions at non-zero energy densities and quantum phase transitions in ground states.

\section{Conclusions} \label{sec:conclusions}

This work provides answers to the questions posed in the introduction.
We showed that the peak in the fidelity susceptibilities of eigenstates far above the ground state coincides with the position of the ergodicity-breaking critical point in the quantum sun model.
While this result is not unexpected, it represents to our knowledge its first high-precision numerical demonstration.
The latter can be achieved due to a precise knowledge about the location of the ergodicity-breaking critical point in the quantum sun model.
Another convenient property of the latter model is that the energy dependence of the critical point, i.e., the many-body mobility edge, can also be determined very accurately for almost arbitrary nonzero energy density~\cite{pawlik_sierant_2024}.
Hence, while it is understood that in ground states the peak of fidelity susceptibility signals the quantum critical point~\cite{zanardi_giorda_07, you_gu_2007, gu_10}, we here showed that at finite energy densities this peak signals the ergodicity-breaking critical point. 

We also argued that fading ergodicity, when applicable, provides a common framework to describe the breakdown of the ETH at the ergodicity-breaking critical point~\cite{kliczkowski_vidmar2024}, as well as the emergence of the peak in the fidelity susceptibilities.
In the fading ergodicity regime, we introduced and numerically tested the ansatz for scale-invariant fidelity susceptibilities, while at the critical point where fading ergodicity terminates, we showed that their maximal values increase as squares of the density of states at the corresponding energy.
The scaling of the maximum of fidelity susceptibilities is a signature of maximal quantum chaos, which was argued to be a consequence of the maximal sensitivity of eigenstates to perturbations~\cite{sels_pandey_2020, lim_matriko_24}.

As a side result, we here introduced a new insight on the energy resolution of the fading ergodicity regime.
Namely, while previous work established fading ergodicity as a precursor regime when approaching the ergodicity-breaking critical point by varying the interaction at a fixed energy density~\cite{kliczkowski_vidmar2024}, the results in Fig.~\ref{figS5} reveal a similar behavior when lowering the energy at fixed interaction.
In the future, this perspective may be useful for establishing a quantitative framework for the ETH breakdown when approaching the ground state.
Currently, it is a common belief that the ETH is valid in quantum-chaotic interacting models at any nonzero energy density above the ground state~\cite{dalessio_kafri_16}, however, the quantitative description of the crossover towards the ground state is still far from understood.

\acknowledgements 
We acknowledge discussions with M. Mierzejewski, A. Polkovnikov and D. Sels.
R.Ś. and L.V. acknowledge support from the Slovenian Research and Innovation Agency (ARIS), Research core funding Grants No.~P1-0044, N1-0273, J1-50005 and N1-0369, as well as the Consolidator Grant Boundary-101126364 of the European Research Council (ERC).
We gratefully acknowledge the High Performance Computing Research Infrastructure Eastern Region (HCP RIVR) consortium~\cite{vega1}
and European High Performance Computing Joint Undertaking (EuroHPC JU)~\cite{vega2}
for funding this research by providing computing resources of the HPC system Vega at the Institute of Information sciences~\cite{vega3}.

\appendix

%%%%%%%%%%%%%%%%%%%%%%%%%%%%%%%%%%%%%%%%%%%%%%%%%%%%%%%%%%
\section{Details of numerical calculations}\label{sec:technical}

The quantum sun model studied here contains several sources of randomness, as described in the text below Eq.~(\ref{eq:qsun}).
For example, the thermal quantum dot is modeled by a random matrix, the interactions couple randomly selected spins inside the dot with spins outside the dot, etc.
Therefore, all fidelity susceptibilities are averaged over a large number of Hamiltonian realizations, $N_{\rm r}$.
Specifically, $N_{\rm r} = 10000$ for $L \leq 9$ and $N_{\rm r} \geq 6000, 3000, 1000, 700$ for $L = 10, 11, 12, 13$.  The exception is for the average fidelity susceptibility at $L = 13$, where $N_{\rm r} = 200$, unless stated otherwise.

\begin{figure}[b]
\centering
\includegraphics[width=\columnwidth]{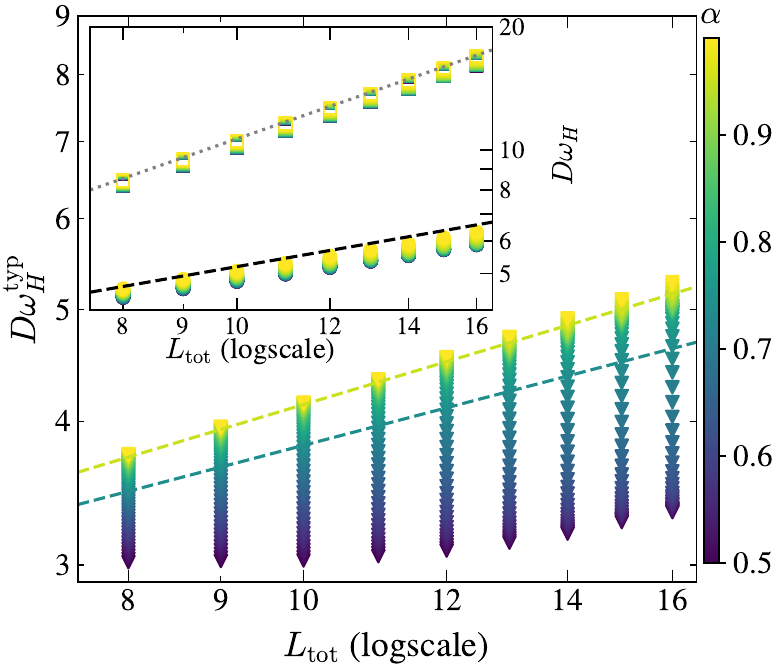}
\caption{
Main panel: 
Finite-size scaling of the scaled typical level spacing, $D\omega_H^{\rm typ}$, for $50\%$ of eigenstates in the middle of the spectrum. 
%Dashed lines indicate the scaling $\propto \sqrt{L_{\rm tot}}$ at the critical point and close to the ETH limit.
Dashed lines are the least-squares fits of $c \sqrt{L_{\rm tot}}$ at the critical point and close to the ETH limit.
Inset: a similar plot for the scaled mean level spacing, $D\omega_H$, calculated from $50\%$ of eigenstates, see filled circles, and from all eigenstates, see empty squares.
Dashed line indicates the scaling $\propto\sqrt{L_{\rm tot}}$, while dotted line indicates the scaling $\propto L_{\rm tot}$.
}
\label{figS1b}
\end{figure}

Next, we compare the scaling of the mean and typical level spacings in the middle of the spectrum.
The mean level spacing is defined as $\omega_H=\expval{\delta_n}_n$, with $\delta_n=E_{n+1}-E_n$, where the average $\langle \cdots\rangle_n$ is taken either over $50\%$ of the states in the middle of the spectrum, or over the entire spectrum.
Simultaneously, the typical level spacing is defined as $\omega_H^{\rm typ}=\exp{\expval{\ln{\delta_n}}_n}$, where the average $\langle \cdots\rangle_n$ is calculated over $50\%$ states in the middle of the spectrum.
In the main panel and in the inset of Fig.~\ref{figS1b}, we show the finite-size scaling of the scaled level spacings $D\omega_H^{\rm typ}$ and $D\omega_H$, respectively.
When considering $50\%$ of the states, we find a qualitatively similar behavior of both $D\omega_H^{\rm typ}$ and $D\omega_H$, i.e., $D\omega_H\propto\sqrt{L_{\rm tot}}$ with $L_{\rm tot}=L+N$, and the prefactor is determined by the coupling $\alpha$.
Hence, in the main text (cf.~Figs.~\ref{fig2} and~\ref{fig5}, and the squares in Fig.~\ref{fig4}), we use the mean level spacing $\omega_H$, and the corresponding density of states $\rho=1/\omega_H$, obtained from $50\%$ of the states in the middle of the spectrum.
We note, in contrast, that the mean level spacing over the entire spectrum scales differently, as $\omega_H \propto L_{\rm tot}/D$, see the upper part of the inset of Fig.~\ref{figS1b}.
This is expected given that the energy bandwidth, i.e., the difference between the maximal and minimal energy, scales linearly with the number of spins, see also Appendix~\ref{sec:mobility_edge}.

%%%%%%%%%%%%%%%%%%%%%%%%%%%%%
%%%%%%%%%%%%%%%%%%%%%%%%%%%%%

%%%%%%%%%%%%%%%%%%%%%%%%%%%%%
\begin{figure*}[t!]
\centering
\includegraphics[width=\textwidth]{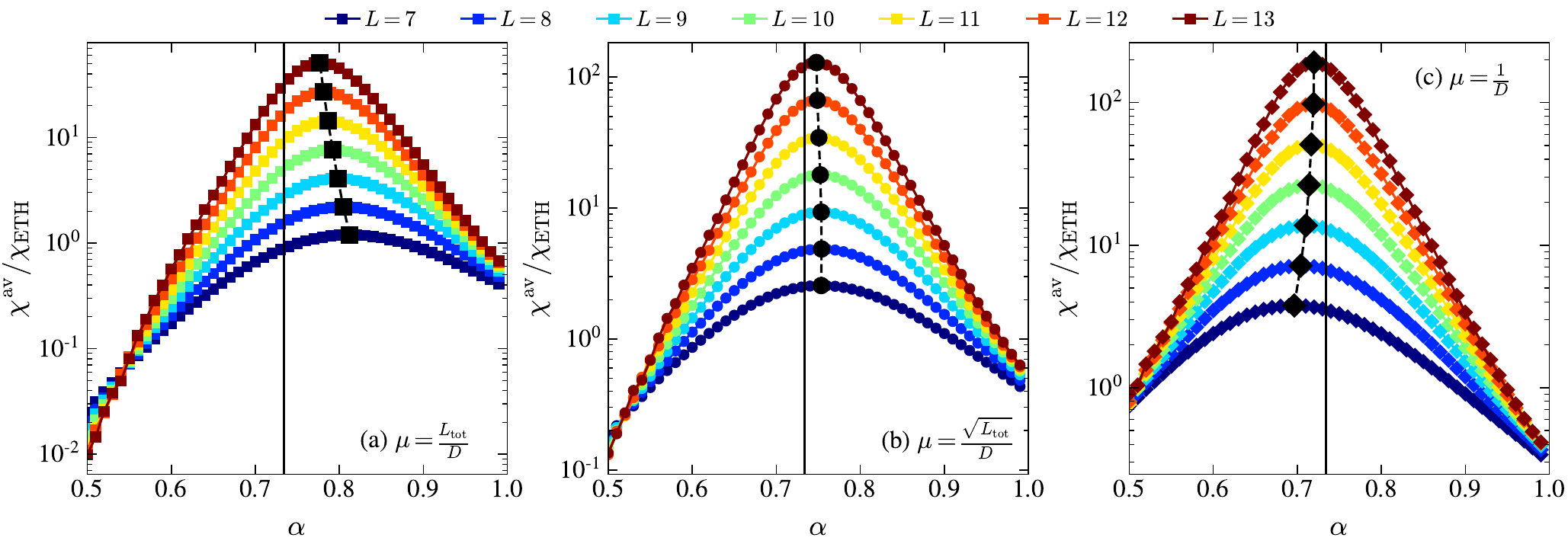}
\caption{
Scaled average fidelity susceptibilities $\chi^\text{av}/\chi_\text{ETH}$ plotted against interaction, $\alpha$. We consider $\hat S_L^z$ as a perturbation.
Each panel corresponds to a different value of $\mu$: (a) $\mu=L_{\rm tot}/D$, (b) $\mu=\sqrt{L_{\rm tot}}/D$, (c) $\mu=1/D$.
Different colors represent results for different system sizes $L$ and the vertical lines mark the critical point, $\alpha_c = 0.734$~\cite{swietek_vidmar_scaling24}.
Large black symbols, obtained from fifth-order polynomial fits to results, mark the positions of the peaks of $\chi^\text{av}/\chi_\text{ETH}$.
Results in panel (b) are identical to those in Fig.~\ref{fig2}(a) from the main text.
}
\label{figS6}
\end{figure*}
%%%%%%%%%%%%%%%%%%%%%%%%%%%%%

Having in mind the scaling of the mean level spacing, let us now turn to the choice of regularization $\mu$ for the average fidelity susceptibility, see Eq.~\eqref{eq:fidelity:average}, which acts as a control parameter to eliminate the effects of (approximate) degeneracies.
However, the choice of this regularization is not unique.
The analysis in Ref.~\cite{kim_polkovnikov_24} proposes that, when the level spacing scales as $\omega_H \propto \sqrt{L_{\rm tot}}/D$, the optimal choice is $\mu \propto L_{\rm tot}/D$.
In this section, we extend the analysis of Fig.~\ref{fig2}(a) in the main text to different values of $\mu$, both for $\mu \geq \omega_H$ and $\mu < \omega_H$, and plot the corresponding $\chi^\text{av}/\chi_\text{ETH}$ in Figs.~\ref{figS6}(a)-\ref{figS6}(c).
All choices of regularization lead to the emergence of a peak in $\chi^\text{av}/\chi_\text{ETH}$.
The precise position of this peak, established from a fifth-order polynomial fit, is sensitive to the value of $\mu$.
Specifically, it approaches the critical point $\alpha_c$ from the right when $\mu > \omega_H$ and from the left when $\mu < \omega_H$.
On the other hand, when $\mu \propto \omega_H$, the peak emerges close to $\alpha_c$ and it exhibits the weakest dependence on $L$, suggesting that this is the most convenient choice.
We hence used $\mu\propto \omega_H$ in the calculations of $\chi^{\rm av}$ in the main text.

When considering properties away from the middle of the spectrum, we defined the microcanonical ensemble average at a target energy ${E}$ with a width $\Delta_\epsilon$ in Eq.~(\ref{eq:fidelity:finiteE:typ}).
For all numerical calculations, the width is fixed at $\Delta_\epsilon = 0.05$, ensuring that the number of states within the microcanonical window, $\mathcal{N}_{\bar E, \Delta\epsilon}$, increases with the system size while remaining a vanishing fraction of the Hilbert space dimension. However, for small system sizes or target energies near spectral edges, if fewer than $10$ states fall within the microcanonical window, the width is adjusted to include exactly $10$ states.
We employ the definition of the microcanonical ensemble to calculate the density of states and the mean level spacing.
Particularly, the density of states at energy ${E}$ (or the corresponding energy density $\epsilon$) is given as $\rho({E})=\mathcal{N}_{E, \Delta\epsilon}/(2\Delta_\epsilon)$, and the mean level spacing at energy $E$ is $\omega_H({E})=1/\rho({E})$.

%%%%%%%%%%%%%%%%%%%%%%%%%%%%%%%%%%%%%%%%%%%%%%%%%%%%%%%%%%
\section{Many-body mobility edge}\label{sec:mobility_edge}

\begin{figure}[!t]
\centering
\includegraphics[width=\columnwidth]{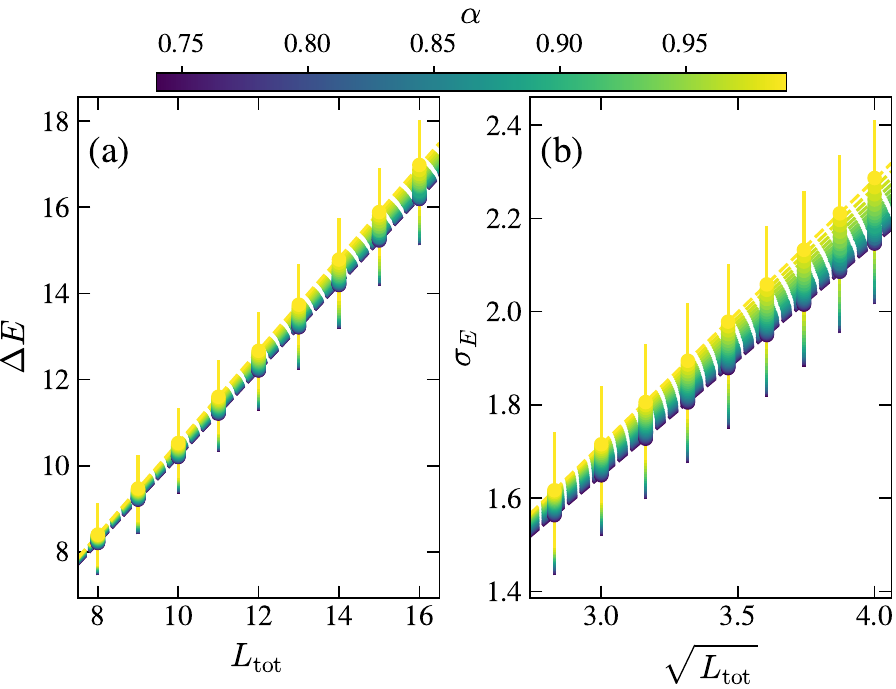}
\caption{
(a) Energy bandwidth $\Delta E$ versus $L_{\rm tot}$, and (b) standard deviation of the energy spectrum $\sigma_E$ versus $\sqrt{L_{\rm tot}}$, where $L_{\rm tot} = L+N$, for different $\alpha$ in the ergodic phase.
Dashed lines correspond to least-squares fits of $f_1(x)=ax+c$ and $f_2(x)=b\sqrt{x}+d$ to $\Delta E$ and $\sigma_E$, respectively. We find that $a \in \qty(0.99, 1.08)$ and $b \in (0.4, 0.5)$.}
\label{figS1}
\end{figure}

We now provide further details about the many-body mobility edge of the quantum sun model.
As demonstrated in Ref.~\cite{pawlik_sierant_2024}, the position of the mobility edge, $\alpha_c(\epsilon)$, can be approximated by the formula
\begin{equation}\label{eq:sm:mobility_edge}
    \alpha_c(\epsilon)=\alpha_c\exp{\frac{a^2(\epsilon-1/2)^2}{4b^2}}\;,
\end{equation}
where $\alpha_c$ on the r.h.s.~of Eq.~\eqref{eq:sm:mobility_edge} denotes the ergodicity-breaking critical point in the middle of the spectrum.
Throughout the paper, we fix the critical interaction strength to $\alpha_c = 0.734$~\cite{swietek_vidmar_scaling24}.
The constants $a$ and $b$ can be determined from the scalings of the energy bandwidth $\Delta E = E_{\rm max} - E_{\rm min}$ and the standard deviation $\sigma_E$ of the energy spectrum, described as
$\Delta E = aL_{\rm tot}$ and $\sigma_E = b\sqrt{L_{\rm tot}}$, see Figs.~\ref{figS1}(a) and \ref{figS1}(b), respectively.
We find that $a \in \qty(0.99, 1.08)$ and $b \in (0.4, 0.5)$.  Thus, we adopt the  midpoints of the intervals as their values, $a = 1.05$ and $b = 0.45$. 

%%%%%%%%%%%%%%%%%%%%%%%%%%%%%
%%%%%%%%%%%%%%%%%%%%%%%%%%%%%

\begin{figure}[b]
\centering
\includegraphics[width=\columnwidth]{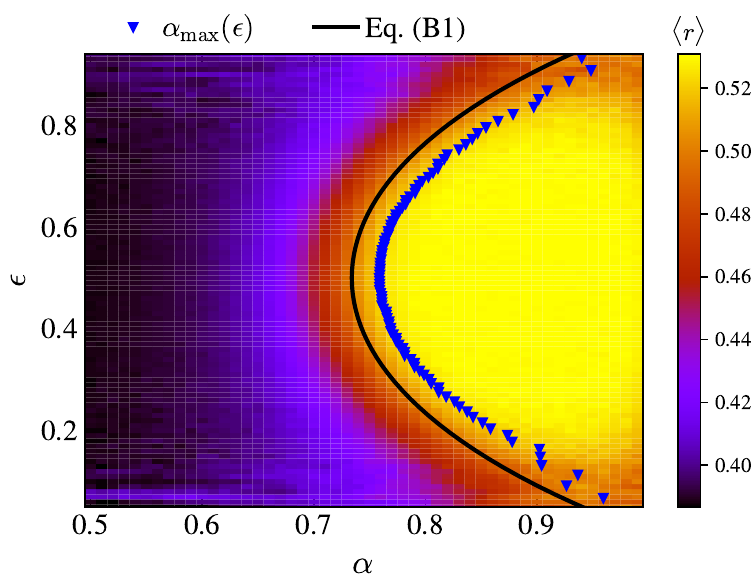}
\caption{
Density plot: the mean gap ratio $\bar{r}$ versus the interaction strength $\alpha$ and the energy density $\epsilon$ for the largest system size $L=13$.
Solid line: the mobility edge prediction $\alpha_c(\epsilon)$ from Eq.~\eqref{eq:sm:mobility_edge}.
Blue triangles: the positions of the maxima, $\alpha_{\rm max}$, of the typical fidelity susceptibility $\chi^\text{typ}(\epsilon)$ from Fig.~\ref{fig3}, at $L=13$.
}
\label{figS2}
\end{figure}

In Fig.~\ref{figS2}, we compare $\alpha_c(\epsilon)$ from Eq.~\eqref{eq:sm:mobility_edge}, see the solid line, with the positions $\alpha_{\rm max}$ of the maxima of typical fidelity susceptibilities $\chi^\text{typ}(\epsilon)$, see the blue triangles.
These results are overlaid on a density plot with interaction strength $\alpha$ and energy density $\epsilon$ on the axes, where different colors represent different values of the mean gap ratio, $\bar r$.
The latter is calculated via $r_n = \max[\delta_n, 1/\delta_n]$ with $\delta_n = E_{n+1} - E_n$, and the average is taken over eigenstates within the same microcanonical window as used for calculating $\chi^\text{typ}(\epsilon)$.
The curves $\alpha_c(\epsilon)$ and $\alpha_{\rm max}(\epsilon)$ closely follow the isoline of $\bar r$ marking its deviation from the GOE prediction $\overline{r} = 0.5307$~\cite{PhysRevB.75.155111, PhysRevLett.110.084101}, therefore suggesting a similar form of the mobility edge.

%%%%%%%%%%%%%%%%%%%%%%%%%%%%%
%%%%%%%%%%%%%%%%%%%%%%%%%%%%%

That said, the values of $\alpha_{\rm max}(\epsilon)$ do not quantitatively match with the values of $\alpha_c(\epsilon)$ predicted by Eq.~\eqref{eq:sm:mobility_edge}.
However, we note that $\alpha_c(\epsilon)$ is the result in the thermodynamic limit, while $\alpha_{\rm max}(\epsilon)$ were established in a finite system at $L=13$.
The position of the ergodicity-breaking critical point is expected to exhibit subleading corrections that vanish in the thermodynamic limit. We model the finite size corrections following the same approach as given in the inset of Fig.~\ref{fig2}(a), and the results in Fig.~\ref{figS3} suggests that it is indeed reasonable to expect $\alpha_{\rm max}(\epsilon)\to \alpha_c(\epsilon)$ in the thermodynamic limit.

%%%%%%%%%%%%%%%%%%%%%%%%%%%%%
\begin{figure*}[t!]
\centering
\includegraphics[width=\textwidth]{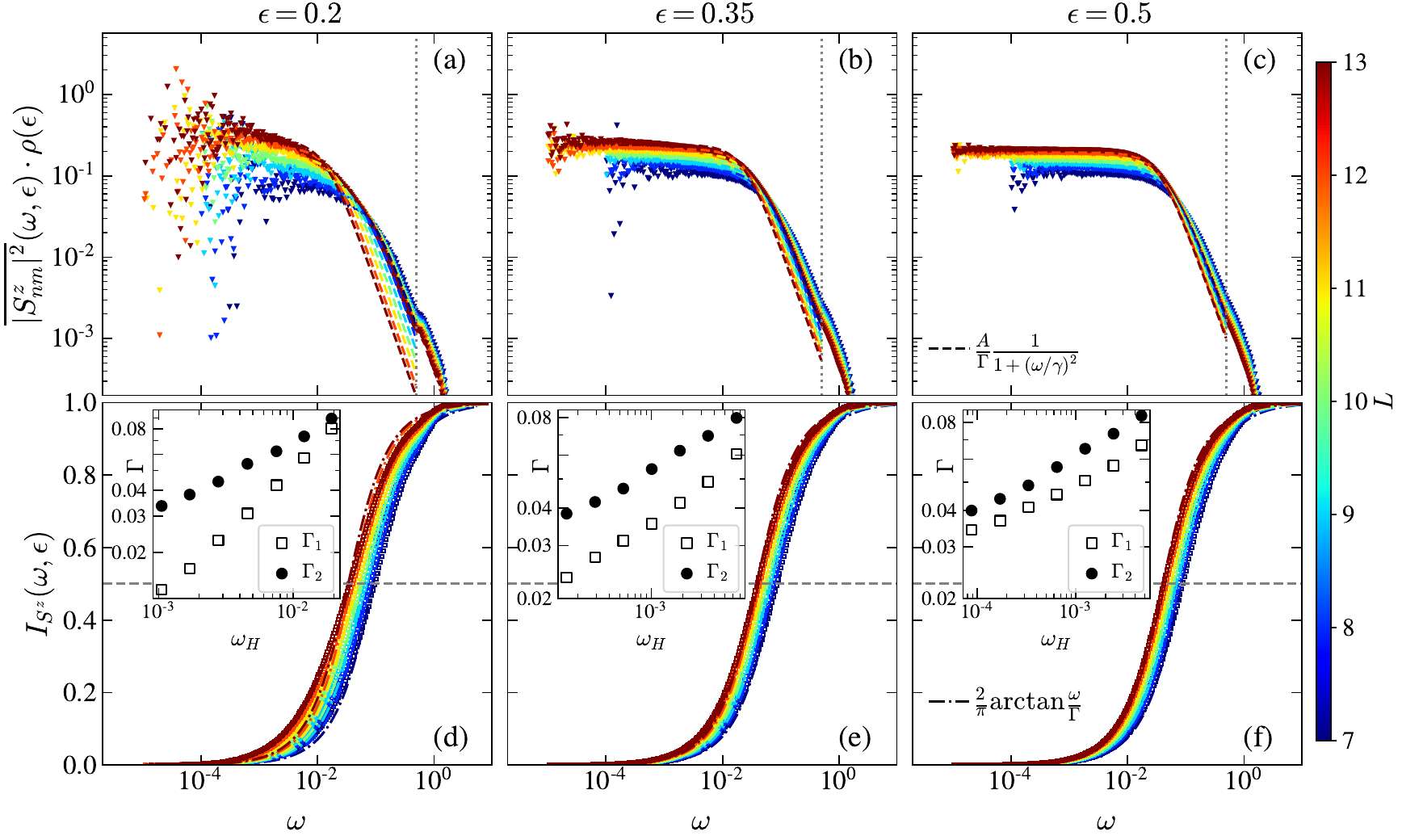}
\caption{(a-c) Scaled coarse grained off-diagonal matrix elements, $\overline{\abs{(S^z_L)_{nm}}^2}(\omega,\epsilon) \cdot \rho(\epsilon)$, and (d-f) integrated spectral functions, $I_{S^z}(\omega,\epsilon)$, for energy densities $\epsilon=0.2,0.35,0.5$, respectively, at $\alpha=0.95$.
Dashed lines in (a-c) correspond to least-squares fits of a Lorentzian function from Eq.~\eqref{eq:sm:lorenz} to numerical results in the interval $\omega\in(\omega_H,0.5)$. 
Dashed-dotted lines in (d-f) mark $2/\pi\arctan(\omega/\Gamma_2)$, with $\Gamma_2$ determined from a simple condition $I_{S^z}(\Gamma_2,\epsilon)=1/2$ from Eq.~\eqref{eq:sm:int_spec_fun:thouless}.
The insets of (d-f) represent the scalings of $\Gamma_1$ (empty black squares) and $\Gamma_2$ (filled black circles) with the levels spacing $\omega_H$.}
\label{figS4}
\end{figure*}
%%%%%%%%%%%%%%%%%%%%%%%%%%%%%

%%%%%%%%%%%%%%%%%%%%%%%%%%%%%%%%%%%%%%%%%%%%%%%%%%%%%%%%%%
\section{Fading ergodicity away from the middle of the spectrum}\label{sec:softening}

Fading ergodicity is understood as the softening of fluctuations of off-diagonal matrix elements in the low-$\omega$ regime, and the diagonal matrix elements, as the system approaches the ergodicity-breaking transition.
The dependence of the fluctuation exponent $\eta$ from Eq.~\eqref{eq:matrix_elements} on the interaction strength $\alpha$ for the quantum sun model in the middle of the spectrum is given by Eq.~\eqref{eq:eta}.
However, as the target energy is varied, the fluctuation exponent $\eta$ is expected to be also a function of the energy density $\epsilon$.
In Fig.~\ref{figS5} of the main text, we plot $\eta$ versus $\alpha$ and $\epsilon$ in the form of a density plot. In this section, we provide a detailed description of the numerical extraction of $\eta$ using Eq.~\eqref{def_Vnm2_eta}.

We consider the coarse-grained off-diagonal matrix elements of the last spin, $\hat{S}^z_L$, within the same microcanonical windows as used for calculating the typical fidelity susceptibility, $\chi^\text{typ}(\epsilon)$. Specifically, we examine
\begin{equation}\label{eq:sm:eta:inf}
    \overline{\abs{(S^z_L)_{nm}}^2}(\omega,\epsilon)=\frac{1}{\mathcal{M}}\sum_{\substack{n\neq m:\\\abs{(\epsilon_n+\epsilon_m)/2-\epsilon}<\Delta_\epsilon\\ \abs{\abs{E_n-E_m}-\omega}<\delta\omega}}\abs{(S^z_L)_{nm}}^2,
\end{equation}
where $\mathcal{M}$ is the number of elements in the sum, while $\delta\omega$ is spaced logarithmically in the range $[0.1/D,\Delta E]$. All numerical results are averaged over $5000,5000,4000,3000,2500,1000$ and $200$ realizations of the Hamiltonian for $L=7,8,9,10,11,12$ and $13$, respectively.
In Figs.~\ref{figS4}(a)-\ref{figS4}(c) we plot these corse-grained off-diagonal matrix elements at $\alpha=0.95$ as functions of energy difference $\omega$ for the selected energy densities $\epsilon$.

The main question when calculating $\eta$ via Eq.~\eqref{def_Vnm2_eta} is at which target energy $\omega$ one should study the fluctuations of the off-diagonal matrix elements.
Ref.~\cite{kliczkowski_vidmar2024} considered $\omega = \sqrt{\omega_H \Gamma}$, where $\Gamma$ is the Thouless energy.
Here, we follow this approach and outline our procedure for the numerical calculation of $\Gamma$ at different energy densities $\epsilon$.

A possible way to extract the Thouless energy $\Gamma$ is to fit the coarse-grained off-diagonal matrix elements (shortly, the spectral function) by a Lorentzian function,
\begin{equation}\label{eq:sm:lorenz}
L(\omega) = \frac{A}{\Gamma_1} \frac{1}{1 + \qty(\omega/\Gamma_1)^2},
\end{equation}
where $\Gamma_1$ is then interpreted as the Thouless energy.
We perform least-squares fits of Eq.~\eqref{eq:sm:lorenz} to the numerical results in the window $\omega_H \leq \omega \leq 0.5$, see the dashed lines in Figs.~\ref{figS4}(a)-\ref{figS4}(c).
At large energy densities, the fit appears reasonable, but significant deviations are observed for $\epsilon \ll 0.5$. Nevertheless, the intermediate regime of $\omega$ closely follows the polynomial decay of $\omega^{-2}$ for all $\epsilon$.
We plot the extracted $\Gamma_1$ versus the level spacing $\omega_H$ in the inset of Fig.~\ref{figS4}(d)-\ref{figS4}(f), see empty black squares.

The observed deviations of the fits from the numerical results also suggest that the values of $\Gamma_1$, at $\epsilon \ll 0.5$, strongly depend on the choice of the fitting window.
Due to these ambiguities, we adopt a different method for extracting the Thouless energy.
Following Ref.~\cite{krajewski_vidmar_22}, we consider the integrated spectral function
\begin{equation}\label{eq:sm:int_spec_fun}
    \tilde{I}_{S^z}(\omega,\epsilon)=\int_0^\omega d\omega\overline{\abs{(S^z_L)_{nm}}^2}(\omega,\epsilon)\;,
\end{equation}
which, for the Lorentzian function, can be calculated analytically as
\begin{equation}
    \int_0^\omega d\omega L(\omega)=A\arctan{\frac{\omega}{\Gamma_2}}\;,
\end{equation}
where $\Gamma_2$ now represents the estimate of the Thouless energy.
Normalizing the integrated spectral function to the interval $[0,1]$,
\begin{equation}
    {I}_{S^z}(\omega,\epsilon)=\frac{\tilde{I}_{S^z}(\omega,\epsilon)-\min{\left[\tilde{I}_{S^z}(\omega,\epsilon)\right]_{\omega\ge0}}}{\max{\left[\tilde{I}_{S^z}(\omega,\epsilon)\right]_{\omega\ge0}}-\min{\left[\tilde{I}_{S^z}(\omega,\epsilon)\right]_{\omega\ge0}}},
\end{equation}
we can extract $\Gamma_2(\epsilon)$ from a simple condition,
\begin{equation}\label{eq:sm:int_spec_fun:thouless}
    {I}_{S^z}(\Gamma_2(\epsilon),\epsilon)=\frac{1}{2}\;,
\end{equation}
where we used that $\max{[\tilde{I}_{S^z}(\omega,\epsilon)]}_{\omega\ge0}=\pi/2$ and $\min{[\tilde{I}_{S^z}(\omega,\epsilon)]}_{\omega\ge0}=0$, while $\arctan{(1)}=\pi/4$.

We calculate the integrated spectral function, ${I}_{S^z}(\omega,\epsilon)$, from the numerical results in Figs.~\ref{figS4}(a)-\ref{figS4}(c), and we find $\Gamma_2$ using the condition from Eq.~\eqref{eq:sm:int_spec_fun:thouless}.
In Figs.~\ref{figS4}(d)-\ref{figS4}(f), we plot ${I}_{S^z}(\omega,\epsilon)$ as points and $2/\pi \arctan (\omega/\Gamma_2)$ as dotted-dashed curves.
As expected, they are similar for large energy densities, while differences are visible for $\epsilon \ll 0.5$.
Nevertheless, $\Gamma_2$ is uniquely determined using this method, as it does not rely on fitting, and hence we use $\Gamma_2$ in our criterion to set the target $\omega$ at which the fluctuations of off-diagonal matrix elements are studied.  
We show $\Gamma_2$ versus $\omega_H$ in the insets of Fig.~\ref{figS4}(d)-\ref{figS4}(f) with filled black circles.
Interestingly, $\Gamma_1$ is consistently smaller than $\Gamma_2$, at least for the studied system sizes. Their scalings with the level spacing, $\omega_H$, are qualitatively similar at large but not at small energy densities $\epsilon$.

\bibliographystyle{biblev1}
\bibliography{references}
\end{document}